# Frequency-Rank Correlations of Rhodopsin Mutations with Tuned Hydropathic Roughness Based on Self-Organized Criticality

J. C. Phillips

Dept. of Physics and Astronomy, Rutgers University, Piscataway, N. J., 08854

Abstract

The behavior of disease-linked mutations of membrane proteins is especially simple in rhodopsin, where they are well-studied, as they are responsible for *retinitis pigmentosa,* RP (retinal degeneration).  Here we show that the frequency of occurrence of single RP mutations is strongly influenced by their posttranslational survival rates, and that this survival correlates well (82%) with a long-range, non-local hydropathic measure of the roughness of the water interfaces of ex-membrane rhodopsin based on self-organized criticality (SOC).  It is speculated that this concept may be generally useful in studying survival rates of many mutated proteins.

**Introduction**

  Rhodopsin (348 amino acids, aa) is the best studied transmembrane protein.  It belongs to the Guanine Protein Coupled Receptor (GPCR) superfamily, the largest family of proteins in the human genome (800 members).  GPCR proteins have characteristic heptad structures, with seven long (25aa), predominantly helical, TransMembrane (TM) interior segments connected by exterior or surface ExtraCellular (EC) and interior CytoPlasmic (CP) loops.  Their amino acid sequences form the largest database for protein-membrane interactions, and they perform a variety of functions: rhodopsin (visual signaling), adrenopsin (stimulative), adenopsin (metabolic), etc. Rhodopsin rods also comprises 95% of the visual mosaic. They provide mechanical support for the longer wave length cone opsins responsible for trichromatic vision in some primates, including humans, which lie at the center of the retinal area.

Our aim here is to test the precision and content of contrasting hydropathicity $\Psi$ scales by exploiting the rapidly growing data base for disease-linked mutations of membrane proteins. Rhodopsin is an excellent test case, because degeneration of the retinal rod mosaic (*retinitis pigmentosa*, RP) is caused by single aa rhodopsin mutations (more than 100 are known). One can regard genetic variations between vertebrate species as a weak kind of mutation. At any amino acid position in human rhodopsin, the set of disease-associated amino acid mutations is strong: it does not show any commonality with the set of weak amino acid mutations present among vertebrate species. The contrast is lessened at the codon level, but even here the average biochemical distance (measured by the Grantham mutational scale) of disease mutations is four times larger than the interspecific (neutral) variation[1].

There are two next steps, predicting the average age of RP blindness for a given mutation of folded rhodopsin, and predicting the frequency of a given RP mutation. Total energy calculations based on five X ray structures of rhodopsin found a highly significant correlation between relative mutational energy changes calculated using the online FoldX algorithm and the average age of night blindness as well as daytime vision loss onset[2]. These energy differences thus correspond to an activation energy for aging. Here we test the ability of hydropathicity scales to predict the mutational frequency; we were led to attempt this in the hope that the rates of mutational transcription and survival could be dominated by specific features of water-protein interactions. This hope is based on the picture of protein translation as dominated by tumbling motions of rougher or smoother globular proteins, with correspondingly larger or smaller water-protein interfacial areas. For homologous proteins differing only by a few % mutations, knowledge of the details of such translation or of the underlying protein folding, are not needed.  Just as only mutational energy differences enter the FoldX correlations, so here only their relative translateability, or relative water-protein interfacial areas, are crucial, and these can be quantified as follows.  Note that the magnitudes of water-protein interfacial areas do not enter the calculations directly.  However, they do reappear later in an interesting way that will be discussed.

**Hydropathic $\Psi$ Scales**

A Kauzmann-like transference hydropathic KD $\Psi$ scale, based on partitioning aa between water and vapor (adjusted to reflect inside/outside preferences and other data), was based on "an amalgam of experimental observations derived from the literature"[3]. Using sliding windows, the authors found that the portions of membrane-bound protein sequences that are located within the lipid bilayer are clearly delineated by large uninterrupted areas on the hydrophobic side of the midpoint line. As such, the membrane-spanning segments of these proteins could be identified.  For decades this chemical transference method served to identify transmembrane segments. Although recently it has been superseded by crystallographic structural studies, we will use the KD $\Psi$ scale here as a benchmark for scaling analysis. There are dozens of similar transference scales, as well as local geometrical scales, which could yield results similar to those of KD.

The KD Ψ scale measures an uncontrolled mixture of short-range and long-range hydropathic interactions. A bioinformatic Ψ scale based on the differential geometry of long-range hydropathic interactions alone has recently emerged, denoted here as the Ψ MZ scale[4,5]. Transference Ψ scales (like KD) imply large (first-order) changes in all protein-solvent interactions, and in effect unfold the topological inside-outside relation between hydrophobic and hydrophilic aa in folded proteins. The MZ Ψ scale is based on the concept that all proteins are the result of evolution, which has placed them near a critical point in almost an equilibrium configuration space. The concept of self-organized criticality (SOC)[6] then suggests that weak long range interactions can be scaled with power-law interactions indexed by fractal (dimensionless) exponents (not energies, distances from a centroid, or surface areas). This model is well suited to describing conformational changes (without unfolding) of globular proteins.

This abstract concept was proven by analyzing the properties of solvent-accessible surface areas (SASA) of 5526 protein segments from the Protein Data Bank as a function of the $C_\alpha$ length $W = 2N+1$. Initially these SASA decrease rapidly with N, but power-law behavior takes over for distant residues with $4 \leq N \leq 17$ ($9 \leq W \leq 35$),

$$d\log(SASA(aa))/d\log N = -\psi(aa) \qquad (1)$$

where $\psi(aa)$ was constant for $N \geq 4$. In other words, the decrease of SASA with increasing segment size asymptotically follows a fractal power law $N^{-\psi}$, with dimensionless exponents $\Psi = \{\psi(aa)\}$ specific to each aa. The Ψ scale established by these exponents can be compared numerically to earlier non-critical hydropathic scales based on simple areas (N = 0, W = 1) or KD transfer energies (see Table I of ref. 5, which lists all 20 amino acid constants for three Ψ scales).

In all the homologous cases we have studied so far, including not only rhodopsin, but also a number of other GPCR signaling proteins[7] as well as lysozyme *c* and many nucleoporin repeat transport proteins[8], the thermodynamically second-order nonlocal MZ Ψ distal scale based on SOC has outperformed first-order local transference hydropathic scales. Here we put these local and nonlocal scales to an even more exacting test, by comparing them with the bioinformatic Grantham scale[9], constructed specifically to measure pair mutational distances. Which scale (if any) will be successful in predicting the frequency of local rhodopsin mutations? How do the scaling results compare with those of mutational energies[2] calculated using the online FoldX algorithm?

**Tuned Hydropathic Roughness**

One can define the nonlocal sliding window average by

$$<\psi(j)W> = \Sigma\ \psi(j + i)/W \qquad (2)$$

with $-N \leq i \leq N$. Sliding window averages are not new to proteins; they are mentioned in the titles/abstracts of 100 + papers. However, they have mainly been studied for small $N \leq 4$, where the improvement they make on predictive accuracies of secondary structures is small (from 65% to 68% for alpha-helical structures in general, or 65% to 70-72% if restricted to homologous aligned sequences[10]). It turned out that significant differences in the hydropathic sliding window average $<\psi W>$ (W = window length) of lysozyme *c* (a non-membrane protein) between species occurred only in certain long (W ~ 15) segments, which could be identified by using $<\psi 3>$ MZ profiles, and these correlate well with evolutionary trends in both enzymatic and antimicrobial properties[5]. These correlations probably arise from dynamical tertiary conformational mechanisms. They are not explicable using other tools, such as the $<\psi 3>$ KD profiles, BLAST sequential similarity or conventional backbone Euclidean geometrical structural superpositions.

Here we will see that hydropathic elastic roughening profiles provide another tunable nonlocal handle on long-range interactions, that could be as useful for the latter as short sequence motifs and crystal structures are for exploring short-range (contact) interactions. Roughening is defined simply in terms of the variances of hydropathic $<\psi W>$ sliding window profiles, where the window length W is tuned to optimize resolution of interprotein (for instance, different species or mutational) differences. The generalized hydrophobicity profiles $<\psi_\alpha(i)W>$ exhibit oscillations, and these oscillations are often quite similar for different species over large parts of a given protein. It was shown[5,8] that occasionally there are systematic differences in $<\psi_{MZ}(i)W>$ over medium length segments, which need not be structurally obvious (unlike the transmembrane segments originally studied by KD). Moreover, systematic differences between these "hidden" hydropathically variable segments can correlate extremely well with protein functionality.

In retrospect these strong MZ correlations could have been expected, as the conformational changes that determine protein functionality involve tertiary long-range forces, with the changes in short-range interactions limited by rigid secondary structures (helices and strands) whose main role is to stabilize the functional units during long-range conformational changes. Increasing W smooths the oscillations of the hydrophobicity profiles and reduces their amplitudes, and yields parameter-free measures of long-range interfacial water-protein roughening. The variance function $\mathscr{A}(W)$ is defined by

$$\mathscr{A}(W) = \sigma^2(W) = \Sigma_i \, (<\psi(i)W(N)> - <<\psi(i)W(N)>>)^2/(M-W) \qquad (3)$$

where the sequence contains M aa, and the average is over the central sites, with N sites at each end excluded.

If we simply use the variance $\mathscr{A}(W)$ relative to an average over the entire unfolded GPCR protein, we obtain good results, but for folded GPCR proteins in membranes sometimes we can do even better: we can average separately over the three transmembrane (TM), extracellular (EC) and cytoplasmic (CP) regions as listed in Uniprot, and calculate the variance by averaging over all three regions with physically distinct background averages $<\psi(i)W(N)>_\alpha$ for each region $\alpha$ = TM, EC or CP. We refer to this variance as the separated roughness form factor

$$\mathscr{R}^*(W) = \sigma^2(W) = \Sigma_\alpha \, \Sigma_i \, (<\psi(i)W(N)> - <<\psi(i)W(N)>_\alpha>)^2/(M-W) \qquad (4)$$

The idea of (4) is that evolution and plastic adaptivity[11] of in-[ex-]situ membrane proteins should be better reflected in species trends of $\mathscr{R}^*(W)$ [$\mathscr{A}(W)$]. One of the advantages of a long-range hydropathic scale is that it is very well suited to separating the in-situ folded α =TM, EC and CP blocks, which nature refines to extrema and either smoothes or roughens (one or the other extremum[12]) in order to optimize the night signaling ability of rhodopsin[11].

Before we proceed further, a comment is in order. The concepts of hydropathic roughness expressed by Eqns. (3) and (4) may seem artificial to some readers. However, on the basis of our earlier studies[5,7,8] of SOC, these concepts seem natural, and they were the first ones tried as a means of explaining frequency/rank correlations. Their success was immediate and it ended the search. It will be interesting to see if similar success can be achieved by another approach not based on SOC, a profound concept that is a remarkable mathematical embodiment of evolutionary optimization[4,6].

**Experimental Data**

Table S1 of ref. 2 lists in column S the total number of reports of data points for each rhodopsin mutation. These numbers may not represent objective mutational frequencies because of subjective reporting bias (the tendency to report lesser known (more "novel") mutations over already better known ones). We address this question in the scaling analysis below. By far the most common mutation is P23H; its dominance has always been mysterious, and in the FOLDX calculations it looks much like all the other strongly destabilized mutations, as does also T58R, the second most common mutation[2]. Among the leading mutations, V137M is a problem, and ref. 2 suggests several reasons why it might not be grouped with the others, including its FoldX instability. We tested this point as well. A subset of the leading RP mutations (R135L, P347L, T17M, V345L,…) is not explained by FoldX mutational energies. They were grouped among 19 "functional residues" in Fig. 5 of ref. 2, and their appearance is explained in terms of post translational modifications. Ref. 2 suggests that these post translational modifications may involve glycosylation. These glycosylated mutations are not a problem for us; our results support the post translational interpretation of ref. 2, not only for the glycosylated subset, but also for the frequency/rank values for all the leading mutations. This suggests a more general explanation for translational effects on frequency of mutations, as distinguished from aging effects, which is discussed below.

We surveyed the first 23 most numerous reported mutations, from 44 to 4 reports, but as discussed below, found most objective correlations only from 44 to 10 double-digit data points each (the first nine mutations). We considered many possible explanations for these frequencies, but the data of ref. 2 are sufficiently extensive to show that only one explanation works well.

**Methods and Results**

The comparative advantage of scaling methods lies in their flexibility, as we can examine both $\mathscr{A}(W)$ and $\mathscr{R}^*(W)$ and tune W to maximize frequency/rank correlations. Here we also have to decide which group of mutations is least affected by subjective factors. Not surprisingly, chemical trends for wild folded membrane GPCR properties are generally optimized[7] with W = 25 (TM length), especially for $\mathscr{R}^*_{MZ}$, which reflects the SOC of the membrane-folded protein interface. However, disease mutations at evolutionarily conserved sites can also alter ex-situ protein translocation. We find the best frequency correlations occur for W = 3 with $\mathscr{R}_{MZ}$, corresponding to the optimized range W = 3 of water-specific protein interactions found in non-membrane (but still globular) proteins like lysozyme[5]. Table 1 shows the correlations of $\mathscr{R}_{MZ}(3)$ variances with frequency for the first nine most common mutations. Excluding V137M significantly improves most correlations. The correlations with the FoldX values are surprisingly good, considering that about half of these mutations were grouped as "functional" [glycosylated] in ref. 2. The correlations with KD Ψ variances and Grantham mutational distances[1,9] are poor.

**Discussion**

It may seem surprising that scaling can decide which mutational frequencies are objective, and which are subjective. Certainly the situation is complex, and Table 2 illustrates how sensitive the results are to choice of variables. The sign of correlation(frequency, $\mathscr{R}_{MZ}(3)$) is constant, and the correlation peaks at N = 9 (corresponding to 9 frequencies (number of data points) between 44 and 10). Even the small change from globular $\mathscr{R}_{MZ}(3)$ to membrane $\mathscr{R}_{MZ}^*(3)$ spoils the correlation pattern: although the maximum magnitude value is reduced only from $|\mathscr{R}_{MZ}(3)|$ = 0.72 to $|\mathscr{R}_{MZ}^*(3)|$ = 0.53 for N = 9, the value of $\mathscr{R}_{MZ}^*(3)$ changes sign between N = 9 and 13. This is not surprising, as $\mathscr{R}_{MZ}^*(3)$ refers to the mutated rhodopsin after insertion and folding in the membrane. Previous analysis[2] had already suggested that the largest effects of mutation on protein aging are probably extra-membrane (posttranslational, preinsertional) for a glycosylated subset of RP functional mutations, but these simulations provide no guide to mutational frequency.

The numbers of data points, the values of the FoldX mutational energies, the Grantham mutational distances dG, and the two roughness variances for folded and unfolded sequences, as derived from the SOC MZ hydropathicity scale for W = 3, are listed for the first 13 mutations (with V137M deleted) in Table 3. The last columns shows $\mathscr{R}_{MZ}(3)$ normalized by its wild value. Note that mutated P23H rhodopsin is significantly smoother (smaller $\mathscr{R}_{MZ}(3)$) even than wild rhodopsin. This suggests that it is more likely to survive post translation and membrane insertion than wild rhodopsin. Smoothing explains why the overall trends in $\mathscr{R}_{MZ}(3)$ correlate so well with the number of data points. Note that this explanation for the strong frequency/rank correlations obtained with $\mathscr{R}_{MZ}(3)$ apply not only to P23H, but to all the disruptive mutations, and do not involve glycosylation of the non-functional residues only[2].

The reader will note in Table 1 that the correlations of roughness with frequencies reverse sign between $\mathscr{R}_{MZ}(3)$ and $\mathscr{R'}_{MZ}(3)$. While the former correlation is superior, the latter is still significant, so this sign reversal should be explained. Clearly it depends in detail on the nature of the rhodopsin fold. As we noted earlier, so long as we are comparing like with like, the nature of this fold need not be known to establish single mutational correlations. However, $\mathscr{R'}_{MZ}(3)$ recognizes implicitly the protein-membrane interactions, while $\mathscr{R}_{MZ}(3)$ does not. Specifically, $<\psi(i)W(N)>_{TM} > <\psi(i)W(N)>_{\alpha}$ for $\alpha$ = EC or CP, the classic application of hydropathicity to protein-membrane interactions[3]. A dramatic example of how this distinction affects structural analysis was given earlier for the effects of high pressure on river and deep sea (high pressure) lamprey rhodopsins, which are encoded by a single gene[13]. The aa of the two species differ at 29 out of 353 sites, and three of these have been identified as responsible for causing a blue shift in the rhodopsin absorption spectra for adaptation to the blue-green photic environment in deep water. This leaves 26 aa replacements to be explained. Structurally 20 out of 171 differences are located in TM regions, and only 9 out of 182 in EC and CP loop sites. The predominance of TM substitutions over loops by a factor of 2.2 is understandable in terms of Euclidean geometry, as the increased deep-water pressure compresses internal pore-confined TM motion more than surface loop motion. However, an even more convincing analysis, with a success factor of 3.5, was obtained with non-Euclidean hydropathic profiles based on $\mathscr{R'}_{MZ}(3)$. For further details see ref. 13.

There is a subtle point here that warrants discussion. Given the limited nature of the rhodopsin mutational frequency/rank data base, the overall agreement between the globular roughness $\mathscr{R}_{MZ}$ with frequency, and its improvement over that obtained with membrane-folded roughness $\mathscr{R}^*_{MZ}$, is impressive. It says that the long-range differential geometry of hydropathic interactions, as averaged over a wide range of 5526 protein segments in crystalline environments, gives good results for rhodopsin translocation, and by inference for the translocation of most other unbound proteins as well. This picture assigns a new significance to protein structures determined crystallographically, and the ability of differential geometry to extract some of the essential features even of unbound proteins that are studied by conformational sampling[14]. It is indeed surprising that such significance should emerge from "only" a study of the evolution of hydropathicity with increasing N [Eqn. (1)], even for 5526 protein segments.

But is this really so surprising? Isn't the really surprising feature the very existence of the log-log SASA linearity (self-organized criticality (SOC)[6]) itself? If we think of what happens to a protein during its translation from transcription through the nucleoporin to its functional site[8], it is clear that optimizing the hydropathic smoothness should be at or near the top of the evolutionary agenda. The MZ scale apparently provides a powerful tool for measuring hydropathic smoothness at long length scales, so if we think of translation as involving globular tertiary conformational tumbling[14] through a predominantly aqueous medium, then this result is less surprising. Also it should not be overlooked that we have still an extra handle on length scales through W, which can always be optimized in studying the effects of small (< 1%) mutational changes on protein properties such as translation.

The success of the present frequency-rank analysis depends on two apparently unrelated yet completely reliable bioinformatic and computational studies (ref. 2 and 4). Such success (82% correlation!) can be taken as an indication of the increasing maturity of biophysics. One can speculate that the scaling and roughening concepts specifically discussed here for rhodopsin may be generally useful in studying survival rates of many prolifically mutated proteins, such as viral proteins. Further support for the consistency of the self-organized critical model for protein folding is provided by bioinformatic self-similar scaling of protein packing[15,16].

| Scale | (First 9) | (First 9 −V137M) |
|---|---|---|
| $\mathscr{P}_{MZ}(3)$ | -0.71 | -0.82 |
| $\mathscr{P}_{MZ}^*(3)$ | 0.53 | 0.62 |
| FoldX  | 0.36 | 0.36 |
| $\mathscr{P}_{KD}(3)$ | 0.10 | 0.17 |
| dGran  | 0.04 | 0.08 |

Table 1. Correlations of different parameters with optimized sets of data points. The Grantham pair mutational distance scale[9,1] is the last parameter.

| Scale | N | 5 | 7 | 9 | 13 | 18 | 23 |
|-------|---|---|---|---|----|----|----|
| frequency-R(MZ3)  |   | -0.63 | -0.66 | -0.71 | -0.66 | -0.41 | -0.31 |
| frequency-R*(MZ3) |   | 0.61  | 0.54  | 0.53  | -0.03 | 0.22  | 0.28  |

Table 2. Variations of $\mathscr{P}_{MZ}(3)$ and $\mathscr{R}_{MZ}(3)$ correlations with N, the numbers of data points (V137M included). When V137M is excluded, the magnitudes of the peak correlation values for $\mathscr{P}_{MZ}(3)$ and $\mathscr{P}_{MZ}*(3)$ increase to 82% and 62%, respectively (see Table 1). The superiority of $\mathscr{P}_{MZ}(3)$ to $\mathscr{R}_{MZ}(3)$ is clear, both in magnitudes, and in consistency of behavior for N ≥ 13 with N ≤ 9.

| Data Pts | Mutat | FOLDX | Dg | $\mathscr{D}_{MZ}^*(3)$ | $\mathscr{D}_{MZ}(3)$ | $\mathscr{D}_{MZ}(3)$/wil | X |
|---:|---|---:|---:|---:|---:|---:|---:|
| 44 | P23H | 15.5 | 77 | 921.0 | 1264.9 | 0.9954 | 4.6 |
| 20 | T58R | 4.9 | 71 | 907.7 | 1269.3 | 0.9989 | 1.1 |
| 16 | R135L | 1.3 | 102 | 919.6 | 1270.2 | 0.9996 | 0.4 |
| 14 | G106R | 1.8 | 125 | 903.7 | 1276.1 | 1.0043 | -4.3 |
| 12 | P347L | 0.3 | 98 | 911.4 | 1271.7 | 1.0008 | -0.8 |
| 12 | T17M | -1.6 | 81 | 907.9 | 1270.7 | 1.0000 | 0 |
| 11 | V345L | -0.4 | 32 | 899.9 | 1273.7 | 1.0024 | -2.4 |
| 10 | G114D | 23.5 | 94 | 911.3 | 1275.4 | 1.0037 | -3.7 |
| 8 | G90D | 6.8 | 94 | 913.2 | 1273.6 | 1.0023 | -2.3 |
| 8 | R135W | 11.6 | 101 | 914.4 | 1268.1 | 0.9980 | 2.0 |
| 8 | D190N | 3.6 | 23 | 938.3 | 1271.6 | 1.0007 | -0.7 |
| 8 | V345M | -0.4 | 21 | 934.3 | 1271.5 | 1.0006 | -0.6 |

Table 3. Collected data for the twelve most frequent mutations (V137M excluded). The numbers of data points and the FoldX mutational energies are from ref. 2, and the Grantham mutational distances (dG) from ref. 9. Here $X = 10^3(1-R_{MZ}(3)/\text{wild})$ measures mutational smoothing relative to wild rhodopsin. The wild value of $\mathscr{R}_{MZ}(3)$ is 1270.7. Because $\mathscr{R}_{MZ}^*(3)$ is measured relative to separated membrane-folded region averages, it is smaller than unfolded $\mathscr{R}_{MZ}(3)$. The fact that good correlations are obtained only for the ex-membrane latter tells us that observed mutational frequencies depend on rhodopsin globules in their posttranslational and pre-membrane insertion configurations.

-